\documentclass[superscriptaddress,onecolumn,secnumarabic,
amssymb,amsmath,nobibnotes,aps,prd,showkeys,showpacs,nofootinbib]{revtex4}

\usepackage[latin1]{inputenc}
\usepackage{graphicx}
\usepackage[english]{babel}

\usepackage{amsmath}
\usepackage{amssymb}
\usepackage{amsfonts}
\usepackage{colordvi}
\usepackage{psfrag}
\usepackage{color}

\begin{document}
\def\cL{{\cal L}}
\def\be{\begin{equation}}
\def\ee{\end{equation}}
\def\bea{\begin{eqnarray}}
\def\eea{\end{eqnarray}}
\def\beq{\begin{eqnarray}}
\def\eeq{\end{eqnarray}}
\def\tr{{\rm tr}\, }
\def\nn{\nonumber \\}
\def\e{{\rm e}}


\def\bef{\begin{figure}}
\def\eef{\end{figure}}
\newcommand{\ans}{ansatz }
\newcommand{\eeqn}{\end{eqnarray}}
\newcommand{\bd}{\begin{displaymath}}
\newcommand{\ed}{\end{displaymath}}
\newcommand{\mat}[4]{\left(\begin{array}{cc}{#1}&{#2}\\{#3}&{#4}
\end{array}\right)}
\newcommand{\matr}[9]{\left(\begin{array}{ccc}{#1}&{#2}&{#3}\\
{#4}&{#5}&{#6}\\{#7}&{#8}&{#9}\end{array}\right)}
\newcommand{\matrr}[6]{\left(\begin{array}{cc}{#1}&{#2}\\
{#3}&{#4}\\{#5}&{#6}\end{array}\right)}
\newcommand{\cvb}[3]{#1^{#2}_{#3}}
\def\lsim{\raise0.3ex\hbox{$\;<$\kern-0.75em\raise-1.1ex
e\hbox{$\sim\;$}}}
\def\gsim{\raise0.3ex\hbox{$\;>$\kern-0.75em\raise-1.1ex
\hbox{$\sim\;$}}}
\def\abs#1{\left| #1\right|}
\def\simlt{\mathrel{\lower2.5pt\vbox{\lineskip=0pt\baselineskip=0pt
           \hbox{$<$}\hbox{$\sim$}}}}
\def\simgt{\mathrel{\lower2.5pt\vbox{\lineskip=0pt\baselineskip=0pt
           \hbox{$>$}\hbox{$\sim$}}}}
\def\unity{{\hbox{1\kern-.8mm l}}}
\newcommand{\eps}{\varepsilon}
\def\ep{\epsilon}
\def\ga{\gamma}
\def\Ga{\Gamma}
\def\om{\omega}
\def\omp{{\omega^\prime}}
\def\Om{\Omega}
\def\la{\lambda}
\def\La{\Lambda}
\def\al{\alpha}
\newcommand{\ov}{\overline}
\renewcommand{\to}{\rightarrow}
\renewcommand{\vec}[1]{\mathbf{#1}}
\newcommand{\vect}[1]{\mbox{\boldmath$#1$}}
\def\tm{{\widetilde{m}}}
\def\mcirc{{\stackrel{o}{m}}}
\newcommand{\Dm}{\Delta m}
\newcommand{\dm}{\varepsilon}
\newcommand{\tanb}{\tan\beta}
\newcommand{\nbar}{\tilde{n}}
\newcommand\PM[1]{\begin{pmatrix}#1\end{pmatrix}}
\newcommand{\up}{\uparrow}
\newcommand{\down}{\downarrow}
\def\omE{\omega_{\rm Ter}}
%

\newcommand{\Dsusy}{{susy \hspace{-9.4pt} \slash}\;}
\newcommand{\DCP}{{CP \hspace{-7.4pt} \slash}\;}
\newcommand{\mc}{\mathcal}
\newcommand{\gr}{\mathbf}
\renewcommand{\to}{\rightarrow}
\newcommand{\gtc}{\mathfrak}
\newcommand{\wh}{\widehat}
\newcommand{\br}{\langle}
\newcommand{\kt}{\rangle}


\def\lsim{\mathrel{\mathop  {\hbox{\lower0.5ex\hbox{$\sim$}
\kern-0.8em\lower-0.7ex\hbox{$<$}}}}}
\def\gsim{\mathrel{\mathop  {\hbox{\lower0.5ex\hbox{$\sim$}
\kern-0.8em\lower-0.7ex\hbox{$>$}}}}}

\def\nn{\\  \nonumber}
\def\de{\partial}
\def\brf{{\mathbf f}}
\def\bbf{\bar{\bf f}}
\def\bF{{\bf F}}
\def\bbF{\bar{\bf F}}
\def\bA{{\mathbf A}}
\def\bB{{\mathbf B}}
\def\bG{{\mathbf G}}
\def\bI{{\mathbf I}}
\def\bM{{\mathbf M}}
\def\bY{{\mathbf Y}}
\def\bX{{\mathbf X}}
\def\bS{{\mathbf S}}
\def\bb{{\mathbf b}}
\def\bh{{\mathbf h}}
\def\bg{{\mathbf g}}
\def\bla{{\mathbf \la}}
\def\bmu{\mathbf m }
\def\by{{\mathbf y}}
\def\bmu{\mbox{\boldmath $\mu$} }
\def\bsig{\mbox{\boldmath $\sigma$} }
\def\bunity{{\mathbf 1}}
\def\cA{{\cal A}}
\def\cB{{\cal B}}
\def\cC{{\cal C}}
\def\cD{{\cal D}}
\def\cF{{\cal F}}
\def\cG{{\cal G}}
\def\cH{{\cal H}}
\def\cI{{\cal I}}
\def\cL{{\cal L}}
\def\cN{{\cal N}}
\def\cM{{\cal M}}
\def\cO{{\cal O}}
\def\cR{{\cal R}}
\def\cS{{\cal S}}
\def\cT{{\cal T}}
\def\eV{{\rm eV}}

\title{Chaotization inside Quantum Black Holes}

\author{Andrea Addazi}

\affiliation{ Dipartimento di Fisica,
 Universit\`a di L'Aquila, 67010 Coppito AQ, Italy}
 
 \affiliation{Laboratori Nazionali del Gran Sasso (INFN), 67010 Assergi AQ, Italy}

\date{\today}

\begin{abstract}

We show how the horizon geometry 
and entropy of
a Semiclassical Black Hole can be reconstructed from a system of 
$N>>1$ horizonless conic singularities with average 
opening angle at the horizon $\langle \Theta \rangle=2\pi$. 
This conclusion is strongly motivated by a generalized Wheeler-De Witt equation 
for quantum black holes. 
We will argument how infalling information will be inevitably chaotized 
in these systems. 
A part of the initial probability density will be trapped inside 
the system, in back and forth scatterings among 
conic singularities, for a characteristic time close to the Semiclassical BH life-time.
Further implications on information paradoxes 
are discussed.



\end{abstract}

\maketitle

\section{Introduction and conclusions}

The information paradox of semiclassical black holes \footnote{ See 
 \cite{Bekenstein,Hawking,Susskind:1993if,'tHooft:1984re,Braunstein:2009my,Almheiri:2012rt}
 for classical references on these subjects. }
 could suggest us
that "Nature abhors real horizons" \footnote{This is a citation from the title of 
paper \cite{Kraus:2015zda}.
} \footnote{Some extensions of general relativity are plagued by inconsistencies 
 at classical level. For example in \cite{Addazi:2014mga}, we have discovered 
geodetic instabilities in a branch of black holes' solutions previously suggested in massive gravity.
}.  However, naked singularities 
seem to be unstable solutions with respect to external electromagnetic, gravitiational 
and matter perturbations \cite{Dotti}. So that, Nature seems also to 
abhor naked singularities! On the other hand, 
numerical simulations of stellar collapses seem to inevitably lead to 
naked singularities  \cite{numerical}. 

Recently, we have suggested that information is chaotized inside realistic black holes,
thought as a system of horizonless geometries 
\cite{Addazi:2015gna,Addazi:2015hpa}. 
This approach could be a step toward the solution of the puzzle mentioned above.  
In this paper, inspired by these our previous ones, we 
suggest that 
the BH quantum state as a superposition of the wave functions of {\it a large number of
conic naked singularities}. We will argument how 
a semiclassical BH solutions are asymptotic limits of 
a $N \rightarrow \infty$ of conic singularities (randomly oriented). 
We will show how the geometry of a BH is effectively recovered
by this horizonless system of conic singularities! \footnote{
This could also have implications in astrophysics.
As discussed in  \cite{Virbhadra:1998dy,Virbhadra:1999nm,Virbhadra:2007kw},
 naked singularities could be detected by virtue of 
 their very peculiar signatures in gravitational lensing measures. 
On the other hand, our systems of conic naked singularities
could be differentiated by semiclassical black holes if their 
"frizzyness is large" enough to be measured in gravitational lensings. }
In other words, the thermodynamical proprieties of semiclassical black holes
are recovered by an ensamble of conic singularities 
except for small correction to Bekenstein-Hawking entropy. 
This conclusion is formally motivated by a generalized Wheeler-De Witt equation for quantum black holes 
\cite{DeWitt:1967yk,Carlip:1993sa,Brustein:2012sa} based on the concept of Wald entropy 
\cite{Wald:1993nt}. An alternative argument based on euclidean path integral was given in \cite{Addazi:2015gna,Addazi:2015hpa} 
and it can also be extended to our new {\it ansatz} considering conic singularities
(see appendix B). 

Let us suppose a thought scattering of a plane wave function on 
a system of N conic singularities
\footnote{
Because of this paper is a part of a special dedicated to Einstein and Bohr, 
we retain appropriate to celebrate these genial theoretical physicists 
with a "gedanken experiment" -that we hope it can provide a "breakthrough" conclusion 
about information fate inside a black hole, or at least it can stimulate interesting counter-arguments against our one. 
}. 
Such a plane wave will be scattered among the conic geometries.
The non-relativistic quantum mechanical scattering problem of an incident 
wave function on a conic space-time can be analytically solved. 
In particular, we will see how the 
scattering amplitude can be expressed 
as a simple combination of Bessel and Henkel functions. 
However, sequential scatterings of a wave function
on a large number of randomly oriented conic singularities
will lead to a chaotization of information. 
In fact the resultant wave function is a 
 superposition of the incident wave function 
with ones diffracted by each "scatterators". 
 This can be thought as a wave function scattering on a {\it quantum Sinai billiard} ! \footnote{
 See 
   \cite{QuantumChaos,QCS,QCS2,QCS3,QCS4}
for useful references in quantum chaos theory. } \footnote{Different applications of chaos theory in black holes' physics 
were suggested in \cite{C0}. }

 Infalling information is highly chaotized inside the space-temporal billiard.
 What one will expect is that the initial probability will be fractioned into
two contributions. In fact, a part of the initial probability density will "escape" 
by the system while a part will remain "trapped" forever in the system 
because of back and fourth scatterings, {\it i.e} for all the system life-time. 
The formation of trapped chaotic saddles inside billiards 
seems inevitable. 
In classical chaotic systems, 
these correspond to surfaces of unstable orbits, 
while in quantum system they correspond to a chaotic superposition 
of unstable wave functions. 
This problem 
is treated with a quantum semiclassical approach in our previous contributions \cite{Addazi:2015gna,Addazi:2015hpa},
reviewed in Appendix C. In this paper we will show formalities of 
 the same problem in Born approximation. 
 
So,  infalling quantum pure states
are fractioned into a "forever" trapped state $|TRAPPED \rangle$
and an emitted one $|B.H. \rangle$ (in form of Bekestein-Hawking radiation):
$$|IN\rangle=c_{1}|B.H.\rangle+c_{2}| TRAPPED \rangle$$
where $c_{1,2}$ are complex coefficients 
depending on the particular configuration of 
conic singularities, 
with 
$$|c_{1}|^{2}=|\langle B.H. | IN \rangle|^{2}$$
$$|c_{2}|^{2}=|\langle TRAPPED | IN \rangle|^{2}$$
$$|c_{1}|^{2}+|c_{2}|^{2}=1$$

Principles of quantum mechanics not allow a transition with $|c_{1}|^{2}=1$,
because of $|B.H.\rangle$ is in a mixed entangled state, while 
$|IN\rangle$ is supposed in a pure one. 
However, a combined state 
of $|B.H.\rangle$ and $|TRAPPED\rangle$ can be a pure one. 
In this case, a transition from a $|IN \rangle$ state to a pure combination 
of two mixed states $|B.H.\rangle$ and $|TRAPPED\rangle$ is
allowed by unitary evolutions. 
During the black hole life-time, $|TRAPPED\rangle$ is not accessible 
to an ideal external observer, so that to reconstruct 
the initial pure state from this one is practically impossible. 
So that,
a quantum mechanical approach describing the unitary evolution of 
 wave functions in time has not sense, in this system.
A wave functions' approach can be substitute by a quantum statistical mechanics' 
approach in terms of density matrices. 

However, let us remark that quantum field theory corrections to 
the non-relativistic approach will ulteriorly favor the chaotization of infalling information. 
Quantum fields' interactions are crucially important in our system.
In fact they will "mediate" a new form of quantum dechoerence induced 
by the non-trivial configuration of the space-time. 
Let us consider the (famous) thought experiment of a Bekenstein-Hawking pair
created near the horizon, one infalling and one tunnelling out. 
Of course, they are entangled and this will lead to the (famous) 
firewall paradox in a semiclassical black hole.
What happen in our space-temporal Sinai billiard?
The infalling particle will be chaotized back and forth 
among asperities and it will start a complicated cascade 
inside. In fact, in non-trivial background (thought as a superposition of gravitons) $\langle G....G \rangle$ , 
infalling particles can inelastically scatter on it:
for example an inelastic scattering of an electron can create 
electromagnetic or hadronic cascade as 
$$e^{-}+\langle G.....G \rangle \rightarrow e^{-}e^{+}e^{-} +\langle G.....G \rangle;\,\,\,\,\,\,\,e^{-}+\langle G.....G \rangle \rightarrow e^{-}q\bar{q} +\langle G.....G \rangle$$
$$e^{-}+\langle G.....G \rangle \rightarrow e^{-}\gamma +\langle G.....G \rangle;\,\,\,\,\,\,\,\,e^{-}+\langle G.....G \rangle \rightarrow e^{-}g +\langle G.....G \rangle;\,\,\,\,....$$
and so on, depending on the particular background structure and local 
CM energy of collision \footnote{For the moment, we only consider standard model interactions. 
However, in presence of non-perturbative interactions induced by exotic instantons,
particles' cascades could also violate B/L numbers 
\cite{Addazi:2014ila,Addazi:2015ata,Addazi:2015rwa,Addazi:2015hka,Addazi:2015eca,Addazi:2015fua,Addazi:2015oba,Addazi:2015goa,Addazi:2015yna,Addazi:2015ewa}
On the other hand, 
new non-local interactions can emerge in the cascade near the effective non-local scale.
See \cite{Addazi:2015dxa,Addazi:2015ppa}  for discussions on these aspects.   }.
Iterating chaotic back and forth scattering and fields' interactions, 
the initial external particle will be no more entangled with one one partner 
but with a very large and chaotic system of particles. 
However, this practically means that such a particle is disentangled.
One can also estimate the entanglement entropy in this system. 
In particular, considering a system of $P$ Bekenstein-Hawking couples
cascading inside the system, they will generate $N>>P$ particles, 
exponential increasing with the number of collisions inside the system. 
Let us suppose for simplicity that a fixed number $N$ of particles 
are produced after $n$ processes, 
in turn producing a rate of $\langle \bar{\nu} \rangle P$ particles for each process. 
In this case one can estimate the entanglement entropy inside the system as
$$S_{e.e}=-{\rm Tr} \rho_{{\rm INSIDE}} \log S_{{\rm INSIDE}}\sim n \log P$$

Another further question regards the fate of such a system, 
considering its (semi) Bekenstein-Hawking evaporation.
The emission of trapped probability density  $\rho(T)$ is approximately described by
$$\frac{d\rho(T)}{dT}\sim -\frac{1}{T^{2}}e^{-\Gamma(T) T }$$
where
 $\rho(T)$ 
is the trapped probability density, dependent by
the number of asperities $N_{s}$ as $\rho \sim N_{s}e^{-\Gamma(T) T}$,
where $\Gamma$ parametrizes the {\it effective average deepness} of 
 asperities (trapping $\rho$); 
 and the number of asperities $N_{s}$ is  in turn dependent by the Black hole mass
 evolution 
  $dM/dT=-1/8\pi T^{2}$. 
As a consequence, the trapped information will be exponentially re-emitted in the environment
for $\Gamma(T) \rightarrow 0$. As a consequence, an S-matrix describing the 
entire black holes' life $\langle {\rm collapse}| S|{\rm complete\, evaporation} \rangle$
is unitary. 

We conclude that these new interpretation of quantum black holes as a large ensamble of conic naked 
singularities seem a viable way-out from information paradoxes, leading to intriguing 
chaotic billiard-like dynamical effects in its interior. However, this approach 
remains incomplete. 
For example, we cannot compute Bekenstein-Hawking entropy from this approach
and the physical interpretation of conic naked singularities remain unknown. 
In other words, an UV completion of our model is still unexplored
\footnote{
It is possible that these conic geometries are sustained by topological defects or exotic non-perturbative configurations. 
For example, supercritical cosmic strings generate conic naked singularities \cite{sclassic,criticalCONIC}.
}.

\section{Quantum Black Holes and wave functions}

A possible reinterpretation of Quantum Black Holes 
is based on the following quantum mechanical postulate:
the quantum state of a black hole is described by a quantum wave function $|\Psi\rangle$.
This implies that the BH entropy is described by a
wave function $\langle S_{W}|\Psi \rangle=\Psi(S_{W})$. 
In Appendix A, we discuss mathematical formalities of this approach, 
with its references \footnote{The formalism that we will use 
was suggested in several papers (see Appendix A). 
Here, we consider a different interpretation of generalized Wheeler-De Witt 
equations. 
 }. 

BH wave functions on entropy representation is described by 
\be \label{Utilde3}
\frac{1}{i}\frac{\partial \Psi(S_{w})}{\partial S_{w}}=\Theta \Psi(S_{w})
\ee
Now, we can perform a semiclassical approximation.
We demand that our wave function will be centered 
into the average value 
$\langle S_{w} \rangle=\frac{1}{4}A_{H}$ 
The equation will take the typical semiclassical form 
\be \label{Utilde3}
\frac{1}{i}\frac{\partial \Psi(S_{w})}{\partial S_{w}}=\Theta_{WKB} \Psi(S_{w})
\ee
where 
\be \label{WKB}
\Theta_{WKB}=\left[ 2\pi-\frac{1}{iC_{1}}\left( S_{w}-\langle S_{w}\rangle \right) + ...\right]
\ee
Clearly, such an approximation can be accepted if 
the BH has a large entropy.
Semiclassical equation has, as usual, a gaussian solution 
\be \label{waveFunction}
\Psi(S_{w})=N_{1}e^{-2\pi i S_{w}}e^{-\frac{1}{2C_{1}}\left(S_{w}-\langle S_{w} \rangle\right)^{2}}
\ee
where $N_{1}$ is an opportune normalization. 
Sol.(\ref{waveFunction}) has a clear interpretation: 
in the semiclassical limit, fluctuations around the "saddle point"
are Gaussian distributions. So that $C_{1}$ is just
\be \label{C1}
C_{1}=2\Delta S_{w}^{2}=2\langle (S_{w}-\langle S_{w}\rangle )^{2}\rangle
\ee 
so that Sol.(\ref{wf2}) can be rewritten as 
\be \label{wf2}
\Psi(S_{w})=N_{1}e^{-2\pi i S_{w}}e^{-\frac{1}{4\Delta S_{w}^{2}}\left(S_{w}-\langle S_{w} \rangle\right)^{2}}
\ee
In the dual Fourier space, one can find out a gaussian distribution also for $\Theta$:
\be \label{wave3}
\tilde{\Psi}(\Theta)=N_{2}e^{i \langle S_{w}\rangle \Theta}e^{-\frac{1}{2C_{2}}\left(\Theta-\langle \Theta \rangle\right)^{2}}
\ee
where $\langle \Theta \rangle=2\pi$. $\Theta$ and $S_{w}$ are conjugated variables
satisfying the indetermination principle $\Delta \Theta \Delta S=\hbar/2$.
(\ref{waveFunction}) is interpreted as wave function for a semiclassical black hole. 

However, the same result can be re-obtained as a superposition of a large number 
number of quantum wave functions $\psi_{\Theta}$ with fixed values of $\Theta$.
Among the infinite samples reproducing (\ref{wf2}), 
a lot of possible wave functions $\psi_{\Theta}$ will have 
$\Theta\neq 2\pi$. But from the geometric point of view, 
a $\psi_{\Theta \neq 2\pi}$ describes an horizonless 
conic singularity. 
Such a conic singularity along the z-axis has a metric
\be \label{metriccone}
ds^{2}=-dt^{2}+dr^{2}+\left(1-\frac{\Psi}{2\pi}\right)^{2}r^{2}d\psi^{2}+dz^{2}
\ee 
where $\Psi$ is the deficit angle, related to the opening angle 
as $\Theta=2\pi-\Psi$. 

Let us suppose a sample 
of N conic singularities described by 
the N entropy variables $S_{w}^{(1)},S_{w}^{(2)},...S_{w}^{N}$,
with corresponding wave functions $\psi_{\Theta_{1}}(S_{w}^{(1)})$,
$\psi_{\Theta_{2}}(S_{w}^{(2)})$,...,$\psi_{\Theta_{N}}(S_{w}^{(N)})$.
For $N>>1$, the central limit theorem will guarantee that 
a Random Variable 
$$S_{w}=\sum_{i=1}^{N}S_{w}^{i}$$
will be distributed as gaussians. 
In order to recover a wave function (\ref{wf2}), we 
have to impose only one condition: 
\be \label{AR}
\langle S_{W} \rangle= \frac{1}{N}\sum_{i=1}^{N}S_{W}^{i}
\ee
corresponding to 
$$\langle \Theta \rangle= \frac{1}{N}\sum_{i=1}^{N}\Theta_{i}$$ in the dual space
\footnote{
Of course, examples of distributions $\psi_{\Theta_{i}}$ avoiding central limit theorem, 
like Chauchy-Lorentz one cannot be considered, or at least they can only be a small fraction 
of distributions in the large ensamble of naked singularities. 
In particular, we remind that for C.L. distributions, moments are undefined. 
So that, all examples of distributions with these kind of pathologies cannot contribute to the formation of 
a black hole. }. 

Clearly, relation (\ref{AR}) are expected to be only an approximated one. 
So that, in principle one could distinguish a semiclassical black hole 
to a "fictious"  one by small deviations by Bekenstein-Hawking entropy, 
{\it i.e} by deviations from the central value of (\ref{wf2}):
\be \label{deviation}
|\Delta S|=| \frac{1}{N}\sum_{i=1}^{N}S_{W}^{i}-\langle S_{W} \rangle \rangle|<< \langle S_{W} \rangle 
\ee

As a consequence, 
a system with a large number of horizonless conic singularities 
can have the same entropy of a Quantum Black Hole,
in semiclassical limit, with small corrections from thermality. Their wave functions 
are not entangled, {\it i.e} their associated metrics have 
to be non-interacting. This approximation is reasonable in semiclassical regime, 
in which gravitational interactions among metrics are strongly suppressed
as well as exchanges of matter entropy among metrics. 

In the following sections, we will study what happen to 
a pure state falling toward a system of N conic singularities.

\section{Non-Relativistic Quantum Scattering }

\subsection{Scattering on a single cone}

Let us consider the Schroedinger equation for a particle,
in a cone geometry \footnote{Perhaps this problem could be found in standard test of advanced quantum mechanics and  non-relativistic quantum scattering theory. I have not found any useful references about this particular problem 
of quantum scattering, so that I have just decided to repeat the exercise in all the details. 
 }.
\be \label{Hamiltonian}
i\frac{\partial}{\partial t}\psi(x)=-\frac{\Delta_{c}}{2m}+A\frac{\delta(r-\bar{r})}{r}
\ee
where $\Delta_{c}$ is the Laplacian in the conical geometry.
For simplicity, we have considered a cone with its axis coincident with the z-axis.
In fact, the radius of the cone boundary is $r=\bar{r}$, 
and it can be encoded in the equation as a 
$\delta$-potential, while $A$ is the dimensional "coupling" of the potential. 

As usually done for this type of problem, 
we can separate the variables
as
\be \label{phi}
\psi(t,x)\sim e^{-i\omega t}\phi_{n}(r)\left( \sin n\nu\theta, \cos n\nu\theta\right)^{T},\,\,\,\,\,n=0,1,2,...
\ee
and defining the adimensional parameter $a=2mA$
and
substituting (\ref{phi}) to (\ref{Hamiltonian}) we obtain 
\be \label{fnr}
\frac{d^{2}\phi_{n}(r)}{dr^{2}}+\frac{1}{r}\frac{d\phi_{n}(r)}{dr}+\left[k_{z}^{2}-\frac{n^{2}\nu^{2}}{r^{2}}-\frac{a}{r}\delta(r-\bar{r})\right]\phi_{n}(r)=0
\ee
We demand as contour conditions 
\be \label{fn}
\phi_{n}(a+o^{+})-\phi_{n}(a+o^{-})=0
\ee
so that we can map such a problem 
to another free-like equation
\be \label{free}
\frac{d^{2}\phi_{n}(r)}{dr^{2}}+\frac{1}{r}\frac{d\phi_{n}(r)}{dr}+\left(k_{z}^{2}-\frac{n^{2}\nu^{2}}{r^{2}}\right)f_{n}(r)=0
\ee
This equation can be also rewritten as 
\be \label{free}
\frac{d^{2}u_{n}(r)}{dr^{2}}+\left(k_{z}^{2}-\frac{n^{2}\nu^{2}}{r^{2}}\right)u_{n}(r)=0
\ee
where $u_{n}=r\phi_{n}$ and $k_{z}^{2}$.

The solution (regular) corresponding to the continuous part of the spectrum is 
\be \label{fnap}
\phi_{n}(r)=c_{n}^{0}J_{n \nu}(k_{z} r),\,\,\,\,\,\,r<\bar{r}
\ee
\be \label{fnam}
\phi_{n}(r)=c_{n}^{-}(k_{z})H_{n\nu}^{-}(k_{z} r)-c^{+}(k_{z})H_{n\nu}^{+}(k_{z} r),\,\,\,\,\,\, r>\bar{r}
\ee
These solutions are valid for all values of $a$ in the $\delta$-potential.
Our problem has two matching conditions 
\be \label{match1}
c^{0}_{n}(k_{z})J_{n\nu}(k_{z} \bar{r})=c^{-}_{n}(k_{z})H_{n\nu}^{-}(k_{z} \bar{r})-c^{+}_{n}(k_{z})H_{n\nu}^{+}(k_{z} \bar{r})
\ee
\be \label{match2}
c^{0}_{n}(k_{z})\left[ \frac{a}{k_{z} \bar{r}}J_{n\nu}(k_{z} \bar{r})+J'_{n\nu}(k_{z} \bar{r})\right]=c_{n}^{-}(k_{z})H'^{-}_{n\nu}(k_{z} \bar{r})-c_{n}^{+}(k_{z})H'^{+}_{n\nu}(k_{z} \bar{r})\ee
(prime is the differentiation with respect to the adimensional variable $k_{z} r$).

This problem can be viewed as a scattering one.
The corresponding solution for the S-matrix is 
\be \label{S}
S_{n}(k_{z})=\frac{a J_{n\nu}(k_{z} \bar{r})H_{n\nu}^{-}(k_{z} \bar{r})+2i/\pi}{a J_{n\nu}(k_{z} \bar{r})H_{n\nu}^{+}(k_{z}\bar{r})-2i/\pi}
\ee
related to $f_{n}$ as usual:
$$S_{n}=1+2ik_{z} f_{n}$$
so that 
$$|S_{n}|=1\rightarrow S_{n}=e^{2i\delta_{n}}$$
We also remind as $f_{n}$ is related to this phase $\delta_{n}$:
$$f_{n}=\frac{e^{2i\delta_{l}}-1}{2ik_{z}}=\frac{e^{i\delta_{n}}\sin \delta_{n}}{k_{z}}$$
Let us remind that, as usual, the asymptotic expansion of the radial part of the wave function 
can be written as the sum of the incident plane-wave on the conic geometry and 
the spherical one 
as 
$$\frac{1}{(2\pi)^{3/2}}\left[e^{ik_{z}z}+f(\theta,\phi)\frac{e^{ikr}}{r}\right]$$

\subsection{Non-Relativistic Quantum Scattering on a Space-time Sinai Biliard}

Let us consider a series of scatterings on a large number of N cones, 
disposed with a uniform random distribution of axis. 
Let us suppose a box of $n \times m \times p$ 
cones, $n$ in the x-axis, $m$ in y-axis, $p$ in z-axis
(not necessary disposed as a regular lattice).  
Let us call $\mathcal{N}_{1},\mathcal{N}_{2}$ the sides sited in the xy-planes, 
$\mathcal{M}_{1,2}$ in xz-planes, $\mathcal{P}_{1,2}$ in zy-planes, 
edges of the box of cones. 
Suppose an incident plane wave $\psi_{0}$
on the 2D surface $\mathcal{N}_{1}$,
with $n\times m$ cones:
$n\times m$ conic singularities will diffract the 
incident wave in $n\times m$-components.
We want to evaluate the S-matrix from the in-state 0 to the out-the box one. 
One will expect that a fraction of initial probability density
will escape from the box by the sides $\mathcal{N}_{1,2}\mathcal{M}_{1,2},\mathcal{P}_{1,2}$, 
 another fraction will be trapped "forever" (for a time-life equal to the one of the system) inside the box. 
As a consequence, we have to consider all possible diffraction stories/paths. 
We also have to consider more complicated diffraction paths: the initial wave
can scatter back and forth in the system before going-out.

We can consider the problem as a superposition of 
the initial wave function, assumed as a wave plane, and the diffracted wave functions for each 
conic singularities. In this system, we can label the position of all the conic singularities as 
$(i,j,k)$, where $i=1,..n$, $j=1,..,m$, $k=1,...,p$. 
The total wave function can be written as
\be \label{totalwave}
\phi_{0}+f(\vect{n}_{0},\vect{n}_{111})\frac{e^{ikr_{111}}}{r_{111}}+f(\vect{n}_{0},\vect{n}_{121})\frac{e^{ikr_{121}}}{r_{121}}+...+f(\vect{n}_{0},\vect{n}_{1N1})\frac{e^{ikr_{1N1}}}{r_{1N1}}
\ee
$$+f(\vect{n}_{111},\vect{n}_{121})\frac{e^{ikr_{121}}}{r_{121}}+...+f(\vect{n}_{111},\vect{n}_{1N1})\frac{e^{ikr_{11N}}}{r_{11N}}$$
$$+f(\vect{n}_{111},\vect{n}_{211})\frac{e^{ikr_{211}}}{r_{211}}+f(\vect{n}_{111},\vect{n}_{221})\frac{e^{ikr_{221}}}{r_{221}}+...+f(\vect{n}_{111},\vect{n}_{2M1})\frac{e^{ikr_{2M1}}}{r_{2M1}}+f(\vect{n}_{111},\vect{n}_{212})\frac{e^{ikr_{212}}}{r_{212}}$$
$$+...+f(\vect{n}_{111},\vect{n}_{21P})\frac{e^{ikr_{21P}}}{r_{21P}}+..+f(\vect{n}_{111},\vect{n}_{2MP})\frac{e^{ikr_{2MP}}}{r_{2MP}}+.....$$
where $\vect{n}_{0}$ is the wave versor of the incident plane wave, 
$\vect{n}_{ijk}$ are wave versors of the scattered waves from the conic singularities in positions $ijk$, 
$r_{ijk}$ are radii from positions $ijk$. 

Under this approximation, we can use the transition amplitudes 
of the one scattering problem considered in the previous section. 

The resultant wave function will be a superposition of an infinite series of waves. 
As a consequence, the total wave function will be highly chaotized by the superposition 
of all the scattered waves. 

An S-matrix for one possible diffraction path is
\be \label{Sone}
\langle in|S^{1th-short}|out\rangle=S_{0-111}S_{111-222}S_{222-333}...S_{(n-1)(m-1)(p-1)-(nmp)}
\ee
where $S_{111-222}$ represents the S-matrix for a process  from 
in-state (after a scattering on) $111$ and with an out-state (after a scattering on) $222$.
This formulation  can be consider if and only if the interdistances among singularities 
are much higher than the cones' sizes.

We can write a generic S-matrix for one diffraction path 
as 
\be \label{Sgeneric}
\langle in|S^{Kth}|out \rangle=S_{0-1jk}S_{ijk}S_{i'j'k'}.....S_{(i^{n-1}j^{m-1}k^{p-1})-(i^{n}j^{m}k^{p})}
\ee
 (\ref{Sgeneric})
with conditions 
\be \label{cond1}
i\leq i' \leq i+1
\ee
\be \label{cond2}
j\leq j' \leq j+1
\ee
\be \label{cond1}
k\leq k' \leq k+1
\ee
$$...$$
\be \label{cond1}
i^{n-1}\leq i^{n} \leq i^{n-1}+1
\ee
\be \label{cond2}
j^{m-1}\leq j^{m} \leq j^{m-1}+1
\ee
\be \label{cond1}
k^{p-1}\leq k^{p} \leq k^{p-1}+1
\ee
represent a class of paths 
similar to (\ref{Sone}). 

These class of paths  are "minimal" ones:
 there are not 
back-transitions. 
 "Minimal paths" are $n\times m \times p \times (n-1)$;
 while the number of non-minimal paths 
 will diverge. 

The total S-matrix
is the (infinite) sum on all diffraction paths
\be \label{TOT}
\langle in|S^{OUT}_{n}|out \rangle=\sum_{paths}\langle in |S^{K-th}_{n}|out \rangle
\ee

The S-matrix for one diffraction path cn be written as 
\be \label{explicitSpath}
\left(S^{Kth}\right)_{n}=\prod_{j=first}^{last}\frac{a_{j}J_{n\nu}(k_{j}\bar{r}_{j})H_{n\nu}^{-}(k_{j}\bar{r}_{j})+\frac{2i}{\pi}}{a_{j}J_{n\nu}(k_{j}\bar{r}_{j})H_{n\nu}^{+}(k_{j}\bar{r}_{j})-\frac{2i}{\pi}}
\ee
where the product is performed from the first scattering to the last one, 
and $a_{j},\bar{r}_{j},k_{j}$ depend by the particular j-th conic singularity
($k_{j}$ depends on the direction of the conic axis). 

However let us remark that $S^{OUT} \neq S^{TOT}$: 
a part of the total S-matrix is associated to the trapped part of the wave function. 
Let us call this S-matrix $S^{hidden}$.

On the other hand, (\ref{explicitSpath}) takes only in consideration 
the continuos part of the S matrix, without resonant poles. 

Bound states correspond to poles along negative real energies
on the first Riemann sheet, of the resolvent operator 
$R(z)=(z-H)^{-1}$. In fact, the Hamiltonian is quadratic in momentum 
so that the inversed function $p=\sqrt{2mE}$ 
has a cut on the $[0,+\infty]$ axis, attaching two Riemann sheets. 
However, there will be also other poles at complex energies on the second Riemann
sheet. The two poles correspond to $E_{a}=\mathcal{E}_{a}-i\Gamma_{a}/2$
and $E_{a}^{*}=\mathcal{E}_{a}+i\Gamma_{a}/2$,
{\it. i.e} at the so called scattering resonances 
and anti-scattering resonances.
In particular, $\mathcal{E}_{a}>0$ is the real part of the energy 
while $\Gamma_{a}>0$ corresponds to the resonances' widths. 
 The dependence of the scattering amplitude on energy
is strictly relates to these poles
as
\be \label{fnE}
f({\bf n};E)\simeq f_{c}({\bf n};E)+\sum_{r}\frac{a_{r}({\bf n})}{E-\mathcal{E}_{r}+i\Gamma_{r}/2}
\ee
where $f_{c}$ is a smoothed amplitude corresponding to 
the continuous part of the spectrum
while $a_{r}(\bar{n})$ are the residues of the resonances' poles. 
$f_{c}$ corresponds to the S-matrix (\ref{TOT}) in our case.

As a consequence, the resonants' parts of the amplitude will interfere
among each other and with the non-resonant parts. 

We can also generalize the notion of time delay also to the non-realtivistic 
quantum chaotic mechanics
\be \label{timedelay}
\mathcal{T}(E)=\frac{1}{i}tr\, \frac{d}{dE}ln\, S(E)
\ee
Let us remind that in a theory with $H=H_{0}+V$,
\be \label{ST}
S(E)=1-2\pi \delta(E-H_{0})T(E+i0^{+})
\ee
where $T$ is the transition operator 
\be \label{transitionOp}
T(z)=V+V\frac{1}{z-H}V
\ee
so that (\ref{timedelay})
can also be re-expressed 
in terms of the Hamiltonian 
as 
\be \label{timedelay}
\mathcal{T}(E)=-2Im\, tr\, \left(\frac{1}{E-H+V+i0^{+}}-\frac{1}{E-H_{0}+i0^{+}} \right)=2\pi \Delta D(E)
\ee
where $\Delta D(E)$ is the difference between 
the level densities of the total Hamiltonian 
and the asymptotic free one. 
This relation shows how the time delay is 
related by the resonance spectrum.
Again, $\mathcal{T}(E)$ will diverge for bounds' states, 
so that this is an alternative way to define the bounds' spectrum 
In appendix C, a discussion of chaotic spectrum of resonances
in semiclassical limit are reviewed.

\subsection{Comments on the range of validity of the previous calculations }

The limitations of our approximated calculations shown in the previous section 
are understood and we resume the main relevant ones:

i) these calculations were done under the first order Born approximation.
This approximation can be accepted if the interdistances among the 
conic geometries are much higher than the size of the cones. 
For interdistances comparable to cones' sizes, higher orders' corrections 
have to be considered. 

ii) These calculations are based on simple non-relativistic quantum mechanics.
In relativistic regime, obviously relativistic quantum field theory is the right
framework to use.  

Let us note that:

a) the disposition of conic singularities was assumed completely random.
Otherwise, a chaotization of the quantum wave function is not generically expected:
for a regular disposition of equally oriented conic singularities, one will expect 
a coherent superposition as in regular lattice, having in mind the Bragg's diffraction 
for example.  

b) the problem becomes a trivial one if the wave lenght of the in-coming wave-function 
is comparable to the size of the system. In (\ref{explicitSpath}), this limit corresponds 
to $k_{j} \bar{r}_{j}\simeq 0$. 
 So, we are assuming that $\lambda$ is 
comparable to the size of conic geometries, and that conic geometries have 
sizes comparable each others. 

\subsubsection{Quantum field theories}

In this section we will formally discuss the problem 
of the "box of cones" from a QFT point of view.

Let us return to our "box of cones" gedanken experiment. In this case, a formulation 
of the problem is again simpler than a realistic case:
supposing interdistances much higher than cones' dimensions, 
In this case, we can define a transition amplitude for each cone.
Let us suppose to be interested to calculate the transition amplitude for 
a field configuration $\phi_{0}$ to a field configuration $\phi_{N}$.
$\phi_{0}$ is the initial field configuration defined on a $t_{0}$, 
before entering in the system, while $\phi_{N}$ is a field configuration 
of a time $t_{N}$, corresponding to a an out-going state from the
system.

One example of propagation Path $0-111-222-333-...-nmp-N$ 
\be \label{examplepath}
\langle\phi_{0},t_{0}|\phi_{111,in},t_{111,in}\rangle \langle\phi_{111,in},t_{111,in}|\phi_{111,out},t_{111,out}\rangle   
\langle\phi_{111,out},t_{111,out}|\phi_{222,in},t_{222,in}\rangle
\ee
$$\times \langle\phi_{222,in},t_{222,in}|\phi_{222,out},t_{222,out}\rangle...\langle\phi_{(n-1,m-1,p-1)},t_{(n-1),(m-1),(p-1)}|\phi_{nmp},t_{nmp}\rangle\langle\phi_{n,m,p},t_{n,m,p}|\phi_{N},t_{N}\rangle$$
where $|\phi_{ijk,in},t_{ijk,in}\rangle$ and $|\phi_{ijk,out},t_{ijk,out}\rangle$
are states before and after entering in the conic geometry $ijk$. 
In order to evaluate $\langle \phi_{0},t_{0}|\phi_{nmp},t_{nmp}\rangle$  
 one has to consider all the possible propagation paths 
from the initial position to the $nmp$-th conic singularity.
Orders and summations are the analogous discussed for S-matrices 
in section 3.2. 
We define these amplitudes as
\be \label{MinkoP}
\langle \phi_{ijk},t_{ijk}|\phi_{i'j'k',in},t_{i'j'k',in} \rangle= \int_{\mathcal{M}_{0}}\mathcal{D}\phi e^{iI[\phi]}
\ee
while 
\be \label{MinkoP}
\langle \phi_{ijk,in},t_{ijk,in}|\phi_{ijk,out},t_{ijk,out} \rangle= \int_{\mathcal{M}_{ijk}}\mathcal{D}\phi e^{iI[\phi]}
\ee
where $\mathcal{M}_{0}$ is the Minkowski space-time, while $\mathcal{M}_{ijk}$ is the ijk-cone space-time. 
Again one can easily get that for a large system of naked conic singularities, 
it will exist a class of propagators' paths,
reaching the out state $|\phi_{N}, t_{N} \rangle$ 
only for a time $t_{N}\rightarrow \infty$.
A simple example can be the propagator paths
\be \label{propath}
|\langle \phi_{ijk},t_{ijk}|\phi_{i'j'k'},t_{i'j'k'} \rangle|^{2} |\langle \phi_{ijk},t^{(1)}_{ijk}|\phi_{i'j'k'},t_{i'j'k'}^{(1)} \rangle|^{2} ....|\langle \phi_{ijk},t^{(\infty)}_{ijk}|\phi_{i'j'k'},t_{i'j'k'}^{(\infty)} \rangle|^{2}
\ee
where $t_{ijk}^{\infty}>....>t_{ijk}^{(1)}>t_{ijk}$ and $t_{i'j'k'}^{\infty}>....>t_{i'j'k'}^{(1)}>t_{i'j'k'}$.
This amplitude is non-vanishing in such a system as an infinite sample of other ones. 
We can formally group these propagators in 
a $\langle BOX|BOX \rangle$ propagator, evaluating the probability 
that a field will remain in the box of cones after a time 
larger than the system life-time. 
On the other hand, let call $\langle BOX|OUT \rangle$
and $\langle OUT|OUT \rangle$
the other processes. 

However, considering Standard Model fields (or its extensions), 
interactions among fields have to be 
considered inside the system. 
We can define an expectation value of a generic operator as
\be \label{Oexp}
\langle \mathcal{O} \rangle= \sum_{\{{\rm all\, K-paths}\}}\prod_{\{ijk, {\rm path}\}}\langle ijk| \mathcal{O}^{Kth} |i'j'k' \rangle
\ee
where for example in a $0-111-222-333-....-N$ path 
\be \label{example1}
\prod_{\{ijk, {\rm path}\}}\langle ijk| \mathcal{O}^{Kth} |i'j'k' \rangle=\langle \phi_{0},t_{0}| \mathcal{O}^{Kth}|\phi_{111,in},t_{111,in}\rangle \langle \phi_{111,in},t_{111,in}| \mathcal{O}^{Kth}|\phi_{111,out},t_{111,out}\rangle .....
\ee
and expectation values are evaluated on path integral on conic fixed backgrounds and non trivial geometries connecting cones. 
The formal way to introduce interactions' terms 
is $\mathcal{O}=\mathcal{L}_{int}$. 
For example, one can evaluate the expectation value of a 
$\frac{1}{4}\lambda \phi^{4}$ interaction term following the procedure 
(\ref{Oexp}). 
However, new interaction terms that usually have a zero expectation value  in SM on a Minkowski space-time
can  be non-null in our system. For example, a photon scattering on a non-trivial background
(especially in asperities among cones' connections) can decay into massive particles 
like for example inelastic processes 
as 
$$\gamma +\langle G...G \rangle \rightarrow q\bar{q}+\langle G...G \rangle \rightarrow hadrons + \langle G...G \rangle$$
related to 
\be \label{ope}
\langle A_{\mu}\bar{q}\gamma^{\mu}q \rangle_{Background}\neq 0
\ee
usually avoided by energy-momentum conservation in Minkowski space-time.

In non-relativistic limit, one can consider 
a non-relativistic path integral formulation.
In bracket-notation,
 the propagator from $(x_{0},t_{0})$ to $(x_{1},t_{1})$ is
$$\mathcal{K}(x_{0},t_{0};x,t_{1})=\langle x_{0},t_{0}| x_{1}, t_{1} \rangle$$
 
This will be equivalent to wave functions' formulation considered 
in section 3.1. 
In this case, 
$\langle OUT|OUT \rangle$ will include all possible 
paths leading to the in-coming "ket" $|x_{0},t_{0}\rangle$
 to another "ket" out of the box. 
This a problem is chaotized: 
one has to consider the quantum interference of all paths 
 for all conic geometries. 
 An example among these paths 
 is $0-111-222-333-...-nmp-N$ 
\be \label{examplepath}
\langle x_{0},t_{0}|x_{111,in},t_{111,in}\rangle \langle x_{111,in},t_{111,in}|x_{111,out},t_{111,out}\rangle   
\langle x_{111,out},t_{111,out}|x_{222,in},t_{222,in}\rangle
\ee
$$\times \langle x_{222,in},t_{222,in}|x_{222,out},t_{222,out}\rangle...\langle x_{(n-1,m-1,p-1)},t_{(n-1),(m-1),(p-1)}|x_{nmp},t_{nmp}\rangle$$
where $|x_{ijk,in},t_{ijk,in}\rangle$ and $|x_{ijk,out},t_{ijk,out}\rangle$
are states incoming and outcoming "kets" in the conic geometry $ijk$. 

An example of trapped propagators 
is
\be \label{propath}
|\langle x_{ijk},t_{ijk}|x_{i'j'k'},t_{i'j'k'} \rangle|^{2} |\langle x_{ijk},t^{(1)}_{ijk}|x_{i'j'k'},t_{i'j'k'}^{(1)} \rangle|^{2} ....|\langle x_{ijk},t^{(\infty)}_{ijk}|x_{i'j'k'},t_{i'j'k'}^{(\infty)} \rangle|^{2}
\ee
where $t_{ijk}^{\infty}>....>t_{ijk}^{(1)}>t_{ijk}$ and $t_{i'j'k'}^{\infty}>....>t_{i'j'k'}^{(1)}>t_{i'j'k'}$.
An ensamble of diffraction paths from OUT to BOX states will be chaotically attracted 
into trapped chaotic zones.

\begin{acknowledgments} My work was supported in part by the MIUR research
grant "Theoretical Astroparticle Physics"  
PRIN 2012CPPYP7
and by SdC Progetto speciale Multiasse 
 "La Societ\'a della Conoscenza in Abruzzo" PO FSE Abruzzo $2007-2013$. 

\end{acknowledgments}

\section*{Appendix A: generalized Wheeler-De Witt  equation}

In this section, we report formal details and definitions
of Carlip-Teitelboim approach for BH \cite{Carlip:1993sa}, based
on an extension of the Wheeler-DeWitt equation 
\cite{DeWitt:1967yk}. 
This approach starts from first axioms of quantum mechanics 
applied to Wald formalism \cite{Wald:1993nt}.
The idea was also well developed in \cite{Brustein:2012sa}.

This approach starts from a BH spacetime foliation 
 with constant hypersurfaces $\Sigma$, with a space-like normal versor $n_{a}$. 
Conveniently, one can define a metric on the hypersurface $\Sigma$ as 
$$g_{ab}=h_{\gamma\delta}e^{\gamma}_{a}e^{\delta}_{b}$$
where $e_{a}^{\gamma}$ are the basis vector of the tangent bundle;
$\gamma,\delta$-indices are the so-called induced coordinates. 
Let us consider a section of the hypersurface $\Sigma$
with surface
$$A=-\frac{1}{2}\int_{\Sigma}dS$$
with 
$$dA=a \epsilon \epsilon$$
and $a$ the area element
and $\epsilon=\nabla n$
(we omit indices of $\epsilon,n,\nabla$).

We can conveniently use re-definitions of $A$ and its Lie derivative 
$\mathcal{L}_{n}A$ (on the direction $n$), 
 in term of hypersurface metric and normal verson:
$$A=-\frac{1}{2}\int_{\Sigma}h^{\gamma \delta}n^{a}n^{b}\epsilon_{\gamma a}\epsilon_{\delta b}$$
$$\mathcal{L}_{n}A=\int_{\Sigma}(\mathcal{L}_{n}h^{\gamma \delta})n^{a}n^{b}a\epsilon_{\gamma a}\epsilon_{\delta b}$$

From these one could derive the following final equation \cite{Brustein:2012sa}:
$$\left\{-\frac{1}{2}\frac{\mathcal{L}_{n}A}{A}(t_{0}),\frac{1}{2\pi}S_{w}(t_{1})\right\}=\frac{1}{\sqrt{-g_{00}}}\delta(t_{0}-t_{1})$$
where $S_{w}$ is the Wald Noether charge entropy, formally defined as
$$S_{w}=-2\pi \int_{H}\frac{\partial L}{\partial R_{\gamma a \delta b}}a\epsilon_{\gamma a} \epsilon_{\delta b}$$
integrated on the Horizon surface. 
($L$ has not to be confused with Lie derivative, because it is just the lagrangian density). 

For a stationary space-time metric, the (D-1)-dimensional hypersurface is the product
of the proper time $\tau$ and the (D-2)-dimensional hypersurface $A_{D-1}=\tau A_{D-2}$.
But $\mathcal{L}_{n}A_{D-1}=0$, implying 
$$A_{D-2}\mathcal{L}_{n} \tau+\tau\mathcal{L}_{n}A_{D-2}=0$$
This allows us to express the Lie derivative of the Area $\mathcal{L}_{n}A_{D-2}$
in term of the one of the proper time $\mathcal{L}_{n}\tau$.
However, the so-called opening angle at the horizon is 
$$\Theta=\frac{1}{2}\mathcal{L}_{n}\tau$$
so that we can relate $\Theta$ to the Lie derivative of the area as
$$\Theta=-\frac{\tau}{2}\frac{\mathcal{L}_{n}A_{D-2}}{A_{D-2}}$$
From these relations, we can arrive to a very suggestive one \cite{Brustein:2012sa}
$$\{ \Theta, \frac{1}{2\pi}S_{w}\}=1$$
where $\{...\}$ is the Poisson bracket.
As usually done from classical mechanics to quantum mechanics, 
one could quantize a Black hole 
as $\{...\}\rightarrow \frac{i}{\hbar}[...]$.
This leads to a Schrodinger equation for the wave function $\Psi$
as
$$\frac{\hbar}{i}\delta \Psi(X)+\delta X\Psi(X)=0$$
where $X=(\Theta,T,...)$, and
$$\delta X=[\delta T M+\delta \Theta \frac{1}{2\pi}S_{w}+dC]$$
where $dC=p\delta q$ represents all the possible variation 
of conjugate variables associated to conserved Noether charges; T is the time separation at infinity, 
$M$ is the ADM mass. 
The equation (\ref{Utilde3}) describes $\Psi$ only with respect to $S_{w}$ 
and not other variables. This is a particular case of the one discussed here.

 \section*{Appendix B: euclidean path integral reformulation}
 
 In this section, we will reformulate definition in section 2 in path integral language
 \cite{Addazi:2015gna,Addazi:2015hpa}. 
 
 Let us consider a system of  N conic naked horizonless singularities,
inside a box $\mathcal{M}$, with a surface $\partial \mathcal{M}$.
 This system satisfied the following hypothesis:
 
 I) Partition functions $Z_{I}$ for each metric tensor $g^{I=1,...,N}$
can be formally defined, with 
 N metrics are considered in thermal equilibrium with the box.

 II) The leading order of the total partition 
function $Z_{TOT}$ is the product of each single partition functions, as
\be \label{ZTOT}
Z_{TOT}= \prod_{I=1}^{N} Z_{I}
\ee
This approximation can be reasonably trusted if 
intergeometries' interactions are negligible 
(with respect to the temperature scale of the box).

III) The total average partition function will be 
\be \label{ZTOTFORM}
\langle Z_{TOT}\rangle=e^{-\frac{\beta^{2}}{16\pi}-\frac{\sigma_{\beta}^{2}}{16\pi}}=Z_{E}e^{\frac{-\sigma_{\beta}^{2}}{16\pi}}
\ee
where $Z_{E}$ is the semiclassical euclidean partition function of a semiclassical black hole, 
$\sigma_{\beta}$ the variance of $\beta$-variable. 
This leads to an entropy 
\be \label{lSr}
\langle S \rangle=\frac{\beta^{2}}{16\pi}+\frac{\sigma_{\beta}^{2}}{16\pi}
\ee

  \section*{Appendix C: semiclassical chaotic scattering}
In this section, we review some aspects of semiclassical chaotic scattering 
considered in our previous papers  \cite{Addazi:2015gna,Addazi:2015hpa}.

The semiclassical propagator
can be written as
\be \label{K}
\mathcal{K}_{WKB}(\vect{r},\vect{r}_{0},t)\simeq \sum_{n}\mathcal{A}_{n}(\vect{r},\vect{r}_{0},t)e^{\frac{i}{\hbar}I_{n}}
\ee
summed on all over the classical n-orbits inside our billiard; 
 amplitudes $\mathcal{A}_{n}$
are defined as 
\be \label{An}
\mathcal{A}_{n}(\vect{r},\vect{r}_{0},t)=\frac{1}{(2\pi i \hbar)^{\nu/2}}\sqrt{|det[\partial_{\vect{r}_{0}}\partial_{\vect{r}_{0}}I_{n}[\vect{r},\vect{r}_{0},t]]|}e^{-\frac{i\pi h_{n}}{2}}
\ee
($h_{n}$ counts for the the number of conjugate points along the n-th orbit). 
The amplitude is related to Lyapunov exponents as
\be  \label{stable}
|\mathcal{A}_{n}|\sim |t|^{-\nu/2}
\ee
on stable orbits
\be \label{unstable}
|\mathcal{A}_{n}|\sim exp\left(-\frac{1}{2}\sum_{\lambda_{k}>0}\lambda_{k}t \right)
\ee
on unstable ones.

In our chaotic system, we expect
many resonances.
A spectrum of resonances called 
Pollicott-Ruelle ones characterizes the
chaotic dynamics of our billiard. 
As a consequence, 
transitions or survival probabilities are averaged 
over the large number 
resonances.  

So that, a wavepacket $\psi_{0}$ 
in a region $R$  ($\nu$-dimensional space)
has a quantum survival probability
\be \label{qsp}
\mathcal{P}(t)={\rm tr} \mathcal{I}_{V}({\bf r})e^{-\frac{iHt}{\hbar}}\rho_{0}e^{+\frac{iHt}{\hbar}}
\ee
where the initial density 
matrix $\rho_{0}=|\psi_{0}\rangle \langle \psi_{0}|$,
 $\mathcal{I}_{V}$ is zero for resonances ${\bf r}$ out of the region $V$,
and${\bf 1}$ for resonances 
into $V$. 
In semiclassical approximation, 
the (survival) probability is
\be \label{sps}
\mathcal{P}(t)\simeq \int \frac{d\Gamma_{ph}}{(2\pi \hbar)^{f}}\mathcal{I}_{D}e^{{\bf L}_{cl}t}\tilde{\rho}_{0}+O(\hbar^{-\nu+1})
\ee
$$+\frac{1}{\pi \hbar}\int dE\sum_{e}\sum_{r}\frac{\cos\left(r\frac{S_{e}}{\hbar}-r\frac{\pi}{2}{\bf m}_{e} \right)}{\sqrt{|det({\bf m}_{e}^{r}-{\bf 1})|}}\int_{e}\mathcal{I}_{D}\tilde{\rho}_{0}{\rm Exp}\{{\bf L}_{cl}t\}dt+O(\hbar^{0})$$
where $d\Gamma_{ph}=d\vect{p}d\vect{r}$ is the phase space infinitesimal volume
  and the sum is on all the periodic orbits as mentioned above (primary elementary periodic orbits are labelled in $e$ while the 
number of their repetitions $r$);
${\bf m}_{e}$ is the Maslov index, 
$S_{e}(E)=\int \vect{p}\cdot d\vect{r}$, $\tau_{e}=\int_{E}S_{e}(E)$,
$\mathcal{M}$ is the Poincar\'e map in the 
neighborhood of the r-orbit
(it is a $(2\nu-2)\times (2\nu-2)$ matrix);
${\bf L}_{cl}$ is the Liouvillian operator
 defined  as 
 ${\bf L}_{cl}=\{H_{cl},...\}_{Poisson}$; $\tilde{\rho}_{0}$
 is the Wigner transform 
 of the (initial) density state.
 In particular, 
 ${\bf L}_{cl}$
 defines the Pollicott-Ruelle peaks mentioned above: 
\be\label{PR}
 {\bf L}_{cl}\phi_{m}=\{{\bf H}_{cl},\phi_{m} \}_{Poisson}=\lambda_{m}\phi_{m}
 \ee
 where eigenvalues $\lambda_{m}$ are complex ones and they correspond to P.R. spectrum, 
where eigenstates $\phi_{m}$ are an ortonormal basis composed of Gelfald-Schwartz distributions.
 $Re(\lambda_{m})\leq 0$,
 correspond to an ensamble bounded periodic orbits;
while $Im(\lambda_{n})$ correspond to
decays and instabilities in the system.
Despite of the complicated form of (\ref{sps})
the survival probability 
has a leading order 
 $\mathcal{P}(t)\sim e^{-\gamma(E)t}$, where 
 $\gamma(E)$ is the classical escape probability (from the system). 
This leading order can be obtained by the 0th order of the expansion 
\be \label{exp}
\mathcal{P}(t)\simeq \int \sum_{m}\langle \mathcal{I}_{V}|\phi_{m}(E)\rangle \langle \tilde{\phi}_{m}(E)|e^{\lambda_{m}(E)t}| \phi_{m}(E)\rangle \langle \tilde{\phi}_{m}(E)|\tilde{\rho}_{0}\rangle 
\ee

\end{document}